\documentclass [12pt]{article}
\usepackage{times}
\usepackage{epsfig}
\usepackage{amsmath}
\begin{document}

\newcommand{\Cerenkov}{\v{C}erenkov}
\newcommand{\elp}{\mbox{$\mathrm{e}^+$}}
\newcommand{\elm}{\mbox{$\mathrm{e}^-$}}
\newcommand{\red}{\mbox{$\mathrm{Re}(\delta)$}}
\newcommand{\imd}{\mbox{$\mathrm{Im}(\delta)$}}
\newcommand{\ree}{\mbox{$\mathrm{Re}(\epsilon)$}}
\newcommand{\reel}{\mbox{$\mathrm{Re}(\epsilon_{\mathrm L})$}}
\newcommand{\rex}{\mbox{${\mathrm Re}(x)$}}
\newcommand{\imx}{\mbox{${\mathrm Im}(x)$}}
\newcommand{\rexb}{\mbox{${\mathrm Re}(\overline{x})$}}
\newcommand{\imxb}{\mbox{${\mathrm Im}(\overline{x})$}}
\newcommand{\rey}{\mbox{${\mathrm Re}(y)$}}
\newcommand{\imxp}{\mbox{${\mathrm Im}(x_{+})$}}
\newcommand{\rexm}{\mbox{${\mathrm Re}(x_{-})$}}
\newcommand{\eplu}{\mbox{$\mathrm{E}_{+}$}}
\newcommand{\emin}{\mbox{$\mathrm{E}_{-}$}}
\newcommand{\epm}{\mbox{$\mathrm{E}_{\pm}$}}

\newcommand{\net}{\mbox{$\mathrm{\nu}$}}
\newcommand{\netb}{\mbox{$\mathrm{\overline{\nu}}$}}
\newcommand{\xrrp}{\mbox{$R_+$}}
\newcommand{\xrrpb}{\mbox{$\overline{R}_+$}}
\newcommand{\xrrm}{\mbox{$R_-$}}
\newcommand{\xrrmb}{\mbox{$\overline{R}_-$}}

\newcommand{\pip}{\mbox{$\mathrm{\pi^+}$}}
\newcommand{\pim}{\mbox{$\mathrm{\pi^-}$}}
\newcommand{\pin}{\mbox{$\mathrm{\pi^0}$}}
\newcommand{\cs}{\mbox{$\mathrm{cos}$}}
\newcommand{\sn}{\mbox{$\mathrm{sin}$}}
\newcommand{\gl}{\mbox{$\Gamma_{\mathrm{L}}$}}
\newcommand{\gs}{\mbox{$\Gamma_{\mathrm{S}}$}}

\newcommand{\stat}{\mbox{$\mathrm{stat}$}}
\newcommand{\syst}{\mbox{$\mathrm{syst}$}}

\newcommand{\ppb}{\mbox{$\mathrm{p\overline{p}}$}}
\newcommand{\pbar}{\mbox{$\mathrm{\overline{p}}$}}
\newcommand{\kn}{\mbox{$\mathrm{K^{0}}$}}
\newcommand{\knb}{\mbox{$\overline{\mathrm
K}\protect\vphantom{\mathrm K}^0 $}}

\newcommand{\kl}{\mbox{$\mathrm{K_{L}}$}}
\newcommand{\ks}{\mbox{$\mathrm{K_{S}}$}}
\newcommand{\ksl}{\mbox{$\mathrm{K_{S,L}}$}}
\newcommand{\km}{\mbox{$\mathrm{K^{-}}$}}
\newcommand{\kp}{\mbox{$\mathrm{K^{+}}$}}
\newcommand{\kpm}{\mbox{$\mathrm{K^{\pm}}$}}
\newcommand{\np}{\mbox{$\mathrm{\pi^{-}}$}}
\newcommand{\pp}{\mbox{$\mathrm{\pi^{+}}$}}
\newcommand{\pn}{\mbox{$\mathrm{\pi^{0}}$}}
\newcommand{\ppm}{\mbox{$\mathrm{\pi^{\pm}}$}}
\newcommand{\pmp}{\mbox{$\mathrm{\pi^{\mp}}$}}
\newcommand{\pipi}{\mbox{$\mathrm{\pi^{+}\pi^{-}}$}}
\newcommand{\popo}{\mbox{$\mathrm{\pi^{0}\pi^{0}}$}}
\newcommand{\rrm}{\mbox{$R^-(t)$}}
\newcommand{\rrpq}{\mbox{$\overline{R}{}^+(t)$}}
\newcommand{\ita}{\mbox{$\eta_{+-}$}}
\newcommand{\mita}{\mbox{$|\eta_{+-}|$}}
\newcommand{\itaoo}{\mbox{$\eta_{00}$}}
\newcommand{\itaooo}{\mbox{$\eta_{000}$}}
\newcommand{\itapmo}{\mbox{$\eta_{+-0}$}}
\newcommand{\ts}{\mbox{$\tau_{\mathrm S}$}}
\newcommand{\tsl}{\mbox{$\tau_{\mathrm S,L}$}}
\newcommand{\ta}{\tau}
\newcommand{\tl}{\mathrm{\tau_{L}}}
\newcommand{\dm}{\mbox{$\Delta m$}}
\newcommand{\dl}{\mbox{$\delta_{l}$}}
\newcommand{\dCPT}{\mbox{$\delta_{CPT}$}}
\newcommand{\asm}{\mbox{$\mathrm{A}_{+-}$}}
\newcommand{\dsdq}{\mbox{$\Delta S \! = \! \Delta Q$}}
\renewcommand{\d}{\mbox{$\delta$}}
\newcommand{\D}{\mbox{$\Delta$}}
\newcommand{\e}{\mbox{$\epsilon$}}
\newcommand{\es}{\mbox{$\epsilon_{\mathrm{S}}$}}

\newcommand{\lS}{\mbox{$\lambda_{\mathrm{S}}$}}
\newcommand{\lL}{\mbox{$\lambda_{\mathrm{L}}$}}

\newcommand{\est}{\mbox{$\tilde{\epsilon}_{\mathrm{S}}$}}
\newcommand{\el}{\mbox{$\epsilon_{\mathrm{L}}$}}
\newcommand{\elt}{\mbox{$\tilde{\epsilon}_{\mathrm{L}}$}}

\newcommand{\et}{\mbox{$\varepsilon_{T}$}}
\newcommand{\ppp}{\mbox{$\mathrm{\pi^{+}\pi^{-}\pi^{0}}$}}
\newcommand{\itp}{\mbox{$\eta_{+-0}$}}
\newcommand{\pnpn}{\mbox{$\mathrm{\pi^{0}\pi^{0}}$}}
\newcommand{\pen}{\mbox{$\mathrm{\pi e \nu}$}}
\newcommand{\fsw}{\mbox{$\phi_{sw}$}}
\newcommand{\fan}{\mbox{$\phi_{00}$}}
\newcommand{\fas}{\mbox{$\phi_{+-}$}}
\newcommand{\rbar}{\mbox{$\mathrm{\overline{R}}$}}
\newcommand{\ko}{\mathrm{K^{0}}}
\newcommand{\kob}{\mathrm{\overline{K}{}^{0}}}
\newcommand{\nnp}{\mbox{$\mathrm{\pi^{-}}$}}
\newcommand{\nora}{\mbox{$\alpha$}}
\newcommand{\mev}{\mbox{$\mathrm{MeV} \! / \! c$}}
\newcommand{\gev}{\mbox{$\mathrm{GeV} \! / \! c$}}
\newcommand{\GeV}{\mbox{$\mathrm{GeV}$}}
\newcommand{\akap}{\mbox{$\widehat{\alpha}$}}
\newcommand{\bkap}{\mbox{$\widehat{\beta}$}}
\newcommand{\gkap}{\mbox{$\widehat{\gamma}$}}
\newcommand{\be}{\begin{equation}}
\newcommand{\ee}{\end{equation}}
\newcommand{\bea}{\begin{eqnarray}}
\newcommand{\eea}{\end{eqnarray}}
\newcommand{\nd}[1]{/\hspace{-0.6em} #1}
\newcommand{\nk}{\noindent}
\newcommand{\nn}{\nonumber}
\newcommand{\vsone}{\vspace{1cm}}
\newcommand{\pr}{\paragraph{}}
\newcommand{\re}{\mathrm{Re}}
\newcommand{\im}{\mathrm{Im}}

\newcommand{\GS}{\mathrm{\Gamma_S}}
\newcommand{\GL}{\mathrm{\Gamma_L}}

\def\a{\alpha}
\def\ba{\tilde{\alpha}}
\def\b{\beta}
\def\bb{\tilde{\beta}}

\begin{flushright}
hep-ph/9812326 \\
CERN-TH/98-392 \\
LPTENS-98-47 \\
\end{flushright}
\vspace{8mm}

\begin{center}
{\large \bf
Violation  of  Time-Reversal Invariance
and CPLEAR Measurements} \\
\end{center}

\vspace*{0.45cm}

\begin{center}
{\bf Luis Alvarez-Gaum\'e~,~
Costas Kounnas$^{\dagger}$~,~ Smaragda Lola~} \\
\vspace*{0.1cm}
{\small CERN Theory Division, CH-1211 Geneva,
Switzerland} \\
\vspace*{0.2 cm}
{\bf  Panagiotis ~Pavlopoulos~}\\
{\small Institut $~$f\"ur Physik, University of Basle CH-4056,\\
and CPLEAR Collaboration, CH-1211 Geneva
Switzerland}
\end{center}

\vspace{0.6 cm}

\begin{center}
{\bf ABSTRACT}
\end{center}

{\small
Motivated by the recent CPLEAR measurement on the time-reversal
non-invariance, we review the situation concerning the
experimental measurements of charge conjugation,
parity violation and time reversibility,
in systems with non-Hermitean Hamiltonians.
This includes in particular neutral
meson systems, like $K^0-\bar{K}^0$, $D^0-\bar{D}^0$ and
$B^0-\bar{B}^0$.  We  discuss
the formalism that describes particle-antiparticle
mixing and time evolution of states,
paying particular emphasis to the
orthogonality conditions
of incoming and outgoing states.
As a result, we confirm that the
CPLEAR experiment makes a direct
measurement of  violation of time-reversal
without any assumption of unitarity and $CPT$-violation.
The asymmetry which signifies $T$-violation,
is found to be independent  of time and decay
processes.}

\vspace*{0.8 cm}

\begin{flushleft}
CERN-TH/98-392 \\
December 1998
\end{flushleft}

\vspace*{0.3 cm}

\noindent
\rule[.1in]{15.5cm}{.002in}
$^{\dagger}$ {\small On leave from Ecole Normale Sup\'erieure,
24 rue Lhommond, F-75231, Paris Cedex 05, France.} \\
{\small E-mail addresses: ~~alvarez@nxth04.cern.ch},$~~$
{\small kounnas@nxth04.cern.ch},$~~$ {\small magda@mail.cern.ch}, \\
\hspace*{3.0 cm}{\small Noulis.Pavlopoulos@cern.ch}

\thispagestyle{empty}

\setcounter{page}{0}
\vfill\eject

\section{Introduction}

Recently, the CPLEAR experiment at CERN, reported the
first direct observation of time-reversal violation
in the neutral kaon
 system~\cite{at_paper}. This observation is
made by comparing the probabilities
of a $\bar{K}^0$ state transforming into a
$K^0$ and vice-versa.
CPLEAR produces initial neutral kaons with defined strangeness from
proton-antiproton annihilations
at rest, via the reactions
\begin{equation*}
 \ppb \longrightarrow \biggl \{
           \begin{array}{l}
           \km \pip K^0  \\
           \kp \pim \bar{K}^0 ~,
    \end{array}
\end{equation*}
and tags the neutral kaon strangeness at the production time
by the charge of the accompanying charged kaon.
Since weak interactions do not conserve
strangeness,  the $K^0$ and $\bar{K}^0$ may subsequently transform
into each-other via oscillations with $\Delta S = 2$.
The final strangeness of the neutral kaon
is then tagged through the semi-leptonic
decays of the type
\bea
K^0 (\bar{K}^0) \rightarrow e^{\pm} \pi^{\mp} \bar{\nu}
({\nu}) ~, \nonumber
\eea
where, a positive (negative) lepton charge is associated with
a $K^0$ ($\bar{K}^0$).

In this way, among other quantities, CPLEAR
also measured the asymmetry
\begin{equation}
\label{at1}
A_T^{exp} =
\frac{R(\bar{K}^0~{(t=0)}
\rightarrow \elp\pim\net ~{(t=\tau)}) - R(K^0~{(t=0)}
\rightarrow\elm\pip\netb ~{(t=\tau)})}
     {R(\bar{K}^0~{(t=0)}
\rightarrow \elp\pim\net~{(t=\tau)}) + R(K^0~{(t=0)}
\rightarrow\elm\pip\netb~{(t=\tau)})} ~,
\end{equation}
which parametrizes
the difference of the probability that an initial
$\bar{K}^0(t_i)$ oscillates to a final $K^0(t_f)$,
from the probability that an initial $K^0(t_i)$
oscillates to a final $\bar{K}^0(t_f)$.
The average value of $A_T^{exp}$  was found over the  time interval
from $1\tau_S$ to $20\tau_S$ (where $\tau_S$ is the lifetime of the
short-lived kaon), to be different than zero
by $4 \sigma$ and this has been interpreted
by CPLEAR as the first direct measurement of time-reversal
non-invariance.

However, doubts have been expressed whether the experiment does
provide
such a direct
evidence for $T$-violation. The basic argument
is that decay processes enter in the observables,
making $CP$-violation manifest. The observed effect is
then attributed to these irreversible
processes, rather than $T$-violation.
It is also argued that this is only
a direct effect of the decaying states
being non-orthogonal.

The aim of this work is to clarify these
points. In order to do so, we
are going to re-discuss
the formalism that describes the particle-antiparticle
mixing and time evolution of states
in the kaon system. Since the
Hamiltonian $H$ of the system
is non-Hermitean, the various masses, widths and eigenstates
have to be  found by using bi-unitary transformations\footnote{
Indeed, there exist unitary matrices $V_L$ and $V_R$ such
that $V_L^\dagger H V_R = H_{diagonal}$.
The form of the two unitary matrices is
found by diagonalizing
the Hermitean combinations $H H^\dagger$ and
$H^\dagger H$, while the physical states
are defined by ``rotations'' of the initial
ones, via the same matrices $V_L$ and $V_R$. }.
This is equivalent to  identifying the
form of the matrices and the eigenstates, by
looking consistently at the correct
orthogonality conditions for the incoming and outgoing states.
The analysis is done in section 2,
where we describe the states in
the vector  space of the system, its dual, as well
as the dual complex space.
In section 3, we are going to show that
the theoretical asymmetry which arises
directly from the {\it definition of  $T$-violation},
is independent  of time and decay
processes.  In section 4, we point out that
this is also true for the
experimental asymmetry that CPLEAR uses,
which differs from the theoretical one
due to the appearance of the semileptonic
decays in the process.
In the same section, we
 show that since the experiment uses a specific
search-channel, rather than summing
over all possible modes, {\it no unitarity
or $CPT$-invariance}
arguments enter in the analysis.
Finally, in section 5 we present a summary of the
basic points and conclude that the
CPLEAR experiment indeed makes a direct
measurement of  $T$-violation.

\section{Definition of states in the incoming ${\bf \rm \cal
H}^{in}$  and outgoing ${\bf \rm \cal  H}^{out}$  dual
spaces}

We denote by ${\bf \rm \cal  H}^{in}$ and ${\bf  \rm \cal H}^{out}$
the Hilbert
space of incoming and outgoing (dual) states,
respectively.
\bea
{\bf \rm \cal  H}^{in}~\equiv~
\left\{~|\Psi_I^{in}>~~,~I=1,2,...,n~\right\}~,~~~~~~~
{\bf \rm \cal H}^{out}~\equiv~
\left\{~<\Psi_I^{out}|~,~~I=1,2,...,n {}~\right\} ~,
\eea
$n$ is the dimension of the space and $|\Psi_I^{in}>$ and
$~<\Psi_I^{out}|$
are the  {\it right}- and {\it left}- eigenstates\footnote{
Technically, we assume that the Hamiltonian $H$
is an $ n \times n$ matrix with $n$ well-defined
left- and right- eigenvectors, to avoid some pathological
cases that are irrelevant in the
$K^0-\bar{K}^0$ system.
.} of the effective
Hamiltonian $H$:
\bea
H~|\Psi_I^{in}> &=&\lambda_I~|\Psi_I^{in}>~~, \nonumber\\
{}~<\Psi_I^{out}|~H &=&<\Psi_I^{out}|~\lambda_I ~.
\eea
In this  basis,  the effective Hamiltonian is diagonal
and can be expressed in the
following form in terms of the incoming and outgoing states:
\bea
H=\sum~~|\Psi_I^{in}> \lambda_I <\Psi_I^{out}|~,
{}~~~{\rm with}
{}~~~<\Psi_I^{out}|\Psi_J^{in}>=\delta_{IJ}~,
\eea
where the unity operator $\bf 1$ takes the usual form:
\bea
{\bf 1}= \sum~|\Psi_J^{in}><\Psi_I^{out}|~.
\eea
Up to this point, we {\it do  not assume} that $H$ is  Hermitean;
$H\ne H^{\dagger}$. This implies that the conjugate states
${~<\Psi_I^{out}|}^{\dagger}$ and ${|\Psi_I^{in}>}^{\dagger}$ are not
isomorphic to their duals:
\bea
|\Psi_I^{out}> & \equiv & {<\Psi_I^{out}|}^{\dagger}  ~~\ne  ~~
|\Psi_I^{in}> ~, \nonumber\\
<\Psi_I^{in}| & \equiv  & ~~
{|\Psi_I^{in}>}^{\dagger} ~~ \ne ~~
|\Psi_I^{out}> .
\eea
The vectors, $|\Psi_I^{out}>$ and $~<\Psi_I^{in}|$ are
eigenstates of the $H^{\dagger}$ operator but they {\it are not
eigenstates} of $H$:
\bea
H^{\dagger}~|\Psi_I^{out}>&=&\lambda^*_I~ |\Psi_I^{out}>~,
\nonumber\\
<\Psi_I^{in}|~H^{\dagger}&=&<\Psi_I^{in}|~\lambda^*_I ~.
\eea
{\it Only if} the effective Hamiltonian is Hermitean,
(i.e. $H=H^{\dagger})$,
the conjugate outgoing states become isomorphic to the incoming
ones, $|\Psi_I^{out}>=|\Psi_I^{in}>$;  in this case the eigenvalues
$\lambda_I=\lambda^*_I$ are real.

When $H\ne H^{\dagger}$, the time  evolution  of the incoming and
outgoing
states $|\Psi_I^{in}(t_i)>~~$ and $|\Psi_I^{out}(t_f)>~$  are
obtained
from $|\Psi_I^{in}>$ and $|\Psi_I^{out}>$, using the
evolution operators $e^{-iHt_i}$ and
$e^{-iH^{\dagger}t_f}$ respectively:
\bea
|\Psi_I^{in}(t_i)> &=&e^{-iHt_i}~|\Psi_I^{in}> ~, \nonumber\\
|\Psi_I^{out}{t_f}>&=&e^{-iH^{\dagger}t_f}~|\Psi_I^{out}> ~.
\eea
{}From the above equations, follows the evolution of the conjugate
states:
\bea
<\Psi_I^{in}(t_i)| &=& <\Psi_I^{in}|~ e^{iH^{\dagger}t_i} ~,
\nonumber\\
<\Psi_I^{out}(t_f)|&=& <\Psi_I^{out}|~e^{iHt_f} ~.
\eea

In view of our later discussion, it is important to stress here that
the
inner products among incoming  and outgoing states {\it do not obey}
the usual orthogonality conditions. Indeed,
\bea
<\Psi_I^{out}|\Psi_J^{out}> \ne \delta_{IJ}~~~~{\rm and}
{}~~~~ <\Psi_I^{in}|\Psi_J^{in}> \ne \delta_{IJ} ~.
\eea
On the other hand, the physical incoming and outgoing eigenstates
obey {\it at all times} the orthogonality conditions
\bea
<\Psi_I^{out}(t_f)|\Psi_J^{in}(t_i)>
=<\Psi_I^{out}|e^{-iH\Delta t}|\Psi_J^{in}>
=e^{-i\lambda_I~\Delta t}~\delta_{IJ} ~.
\eea

We now proceed to discuss particle-antiparticle
mixing in the neutral kaon system.

\section{ Particle-antiparticle mixing in the neutral kaon system}
\noindent

The $K^0, \bar{K}^0$ states are produced under strong interactions
and are strangeness eigenstates.
Moreover, they obey the relations:
\begin{eqnarray}
CP \; |K_0^{in}> & = & |\bar{K}^{in}_0> ~, \nonumber \\
T \; |K_0^{in}> & = & <K_0^{out}|  ~, \nonumber \\
CPT \; |K_0^{in}> & = & <\bar{K}_0^{out}| ~.
\label{EQ1}
\end{eqnarray}

These states are admixtures of the physical
{\it incoming} ($|K_S^{in}>$ and $|K_L^{in}>$) and
{\it outgoing} ($<K_S^{out}|$ and $<K_L^{out}|$) states
of the full Hamiltonian and obey the following orthogonality
conditions:
\bea
<K_L^{out}|K_S^{in}> = 0 ~,~~~~~~ <K_S^{out}|K_L^{in}> = 0 ~,
\nonumber \\
<K_S^{out}|K_S^{in}> = 1 ~,~~~~~~ <K_L^{out}|K_L^{in}> = 1 ~.
\label{ortho}
\eea

The physical states, are the left and right
eigenvectors of the
effective Hamiltonian of the system, $H \equiv
M - i \Gamma /2 $:
\bea
& &  H ~|K_L^{in}>   =  \lambda_L ~|K_L^{in}> ~, \; \; \;
H ~|K_S^{in}> = \lambda_S ~|K_S^{in}>  ~,\nonumber \\
& & <K_L^{out}|~ H  = <K_L^{out}|~\lambda_L ~, \; \; \;
<K_S^{out}|~ H  = <K_S^{out}|~ \lambda_S ~.
\eea
Since $H$ is not Hermitean, this implies in general that the incoming
and outgoing eigenvectors in the $K^0$, $\bar{K}^0$ base
{\it are not related} simply by complex conjugation.

Without loss of generality,
we can express the physical incoming states in terms of
$|K_0^{in}>$ and  $|\bar{K}_0^{in}>$ as:
\begin{eqnarray}
|K_S^{in}> & = & \frac{1}{N_S} \left ( \; (1+\a)|K_0^{in}> + \;
            (1-\a) \; |\bar{K}_0^{in}>  \right ) ~, \nonumber \\
|K_L^{in}> & = & \frac{1}{N_L} \left ( \; (1+\b)|K_0^{in}> - \;
            (1-\b) \; |\bar{K}_0^{in}>  \right ) ~,
\label{EQ2}
\end{eqnarray}
where $\a$ and $\b$ are complex variables
associated with $CP,T$ and $CPT$-violation,
and $N_L, N_S$ are normalization factors
to be discussed below.
Similar relations exist for the dual outgoing states:
\begin{eqnarray}
<K_S^{out}| & = & \frac{1}{\tilde{N}_S} \left ( \;
(1+\ba) <K_0^{out}| + \;
            (1-\ba) \; <\bar{K}_0^{out}|  \right ) ~, \nonumber \\
<K_L^{out}| & = & \frac{1}{\tilde{N}_L} \left ( \;
(1+\bb) < {K}_0^{out}| - \;
            (1-\bb) \; <\bar{K}_0^{out}| \right ) ~.
\label{EQ3}
\end{eqnarray}

The parameters ( $\a, \b$) and ($\ba, \bb$)
that are associated with the
incoming and outgoing states respectively, are not independent
but are related through the orthogonality conditions (eqs.13) valid
for the physical states:
\bea
<K_L^{out} | K_S^{in}> = 0 & \Rightarrow & \bb = -\alpha  ~,
\nonumber \\
<K_S^{out} | K_L^{in}> = 0 & \Rightarrow & \ba = -\beta   ~,
\nonumber \\
<K_S^{out} | K_S^{in}> = 1 & \Rightarrow & N_{S} \tilde{N}_S = 2 (1
- \alpha
\beta)
  ~, \nonumber \\
<K_L^{out} | K_L^{in}> = 1 & \Rightarrow & N_{L} \tilde{N}_L = 2 (1
- \alpha
\beta)  ~.
\eea

The above relations indicate that, while the normalizations
$\tilde{N}_{S,L}$ can be expressed in terms of
$N_{S,L}$, the latter remain unspecified. This
ambiguity however will not affect any measurable
quantity. Thus we can always choose
\bea
N \equiv N_{S}= \tilde{N}_S= N_{L} =\tilde{N}_L =\sqrt{2 (1 - \alpha
\beta)} ~.
\eea

Let us write down for completeness the inverse
transformations that express the $K^0, \bar{K}^0$
states in terms of $K_S$ and $K_L$:
\begin{eqnarray}
|K_0^{in}> & = & \frac{1}{N}
            \left ( \; (1-\b)|K_S^{in}> + \;
            (1-\a) \; |K_L^{in}>  \right ) ~,  \nonumber \\
|\bar{K}_0^{in}> & = &  \frac{1}{N}
            \left ( \; (1+\b)|K_S^{in}> - \;
            (1+\a) \; |K_L^{in}>  \right ) ~,
\label{EQ4}
\end{eqnarray}
and
\begin{eqnarray}
<K_0^{out}| & = & \frac{1}{N}
            \left ( \; (1+\a)<K_S^{out}| + \;
            (1+\b) \; <K_L^{out}| \; \right ) ~,  \nonumber \\
<\bar{K}_0^{out}| & = & \frac{1}{N}
            \left ( \; (1-\a)<K_S^{out}| - \;
            (1-\b) \; <K_L^{out}|  \; \right ) ~.
\label{EQ5}
\end{eqnarray}

In the basis of the states $K_L, K_S$,
$H$ can be expressed in terms of a diagonal  $2 \times  2$ matrix
\bea
H = |K_S^{in}> \lambda_S  < K_S^{out}| +
|K_L^{in}> \lambda_L <K_L^{out}| ~,
\eea
while in  the basis of
$K^0,\bar{K}^0$, $H$ takes the following form:

\bea
H_{ij} =  \frac{1}{2} \left(
\begin{array}{cc}
( \lL + \lS) -
\Delta \lambda \frac{ (\a-\b)}{1-\a\b} &
{}~~~~~~~~~~
\Delta \lambda   \;
\frac{(1 + \a\b)}{1-\a\b}
+ \Delta \lambda \frac{ \a+\b}{1-\a\b} \\
 & \\
\Delta \lambda  \;
\frac{(1 + \a\b)}{1-\a\b}
- \Delta \lambda \frac{\a+\b}{1-\a\b} &
{}~~~~~
( \lL + \lS) +
\Delta \lambda \frac{\a-\b}{1-\a\b}
\end{array}
\right ) ~.
\label{eqH}
\eea

Here,
\[
 \Delta \lambda = \lambda_L - \lambda_S, \; \; \;
\lambda_L = m_L - i \frac{\Gamma_L}{2}, \; \; \;
\lambda_S = m_S - i \frac{\Gamma_S}{2}, \; \; \;
\]
where
$m_{\mathrm S}, m_{\mathrm L}$ are the $\ks, \kl$ masses
and $\Gamma_{\mathrm S},\Gamma_{\mathrm L}$,
the $\ks , \kl$ widths.
 From eq.(\ref{eqH}), we can identify
the $T$-, $CP$- and $CPT$- violating parameters. Indeed:

$\bullet$ Under {\underline {$T$--transformations}},
 $$
<K_0^{out}|H |\bar{K}_0^{in}> \; \leftrightarrow \;
<\bar{K}_0^{out}|H|K_0^{in}> ~,
$$
thus, the off-diagonal elements of $H$
are interchanged.
This indicates that the parameter $\epsilon \equiv (\a+\b)/2$,
which is related to the difference of the
off-diagonal elements of $H$,
measures the magnitude of  the $T$-violation\footnote{
$~2/N^2 \approx 1$, in the linear approximation.}.
\begin{equation}
  \frac{2}{N^2}~\e = \frac {
<K_0^{out}|H|\bar{K}_0^{in}> - <\bar{K}_0^{out}|H|K_0^{in}>
}{2 \; \Delta \lambda} ~ .
\label{epsilon1}
\end{equation}

$\bullet$ Under {\underline {$CPT$--transformations}},
$$
<K_0^{out}|H |{ K}_0^{in}>
\; \leftrightarrow \;   <\bar{K}_0^{out}|H |\bar{K}_0^{in}> ~,
$$
and therefore, the parameter
$\delta \equiv (\a-\b)/2$, related to the
difference of the diagonal elements of $H$,
 measures the magnitude
of $CPT$-violation.
\bea
\frac{2}{N^2}~ \d = \frac{<\bar{K}_0^{out}|H|\bar{K}_0^{in}> -
<K_0^{out}|H|K_0^{in}>}
{2 \; \Delta \lambda}  ~.
\eea

$\bullet$ Under {\underline {$CP$--transformation}},
$$
<K_0^{out}|H |K_0^{in}> \; \leftrightarrow \;
<\bar{K}_0^{out}|H |{\bar
K}_0^{in}> ~,
$$
and simultaneously
$$
<K_0^{out}|H |\bar{K}_0^{in}>\; \leftrightarrow \;    <{\bar
K}_0^{out}|H|K_0^{in}> ~,
$$
thus, {\it both} the diagonal and the off-diagonal elements of
$H$
are interchanged.
Then, the parameters $\a=\e+\d$ and $\b=\e-\d$, usually denoted as
$\epsilon_S$ and $\epsilon_L$, are the ones
which measure the  magnitude of $CP$-violation in the decays of
$\ks$ and $\kl$  respectively.

\section{Direct measurement testing time-reversibility}

The meaning of classical time-reversal invariance is unambiguous.
A system at a
final classical configuration
retraces its way back to some initial configuration by reversing the
velocities.  As a result of time-reversal invariance, initial and
final quantum mechanical  states
are interchanged with identical positions and opposite velocities:
\bea
T~[~<\Psi^{out} (t_f)|\Phi^{in}(t_i)>~]~
=~<\Phi^{out} (t_f)|\Psi^{in}(t_i)> ~.
\eea

In order to test time reversibility, one has to compare
the magnitude of the probability
$|<\Psi^{out} (t_f)|\Phi^{in}(t_i)>|^2$
with that of the time-reversed process
$|<\Phi^{out} (t_f)|\Psi^{in}(t_i)>|^2$.
Any possible difference in the two probabilities will signal
deviations of time-reversibility. In that case, the process is not
equivalent to its
time reversed one, resulting in time-reversal
violation.
In the neutral kaon system,  at a given time $t_i$ one has an initial
 strangeness eigenstate, such that $
|K_0^{in}(t_i)> = |\Psi^{in}(t_i)>$.
At some later time $t_f$, one  finds a final strangeness eigenstate
$<\bar{K}_0^{out}(t_f)| = <\Phi^{out}(t_f)|$.
According to time-reversibility, we may conclude that
the above process should have the same probability with the reversed
one, namely, an initial $|\bar{K}_0^{in}(t_i)>$ to be transformed
into a
final
$<K_0^{out} (t_f)|$. Then, for the kaon system we can write
for the case of time-reversal invariance:
 \begin{equation}
|<\bar{K}_0^{out} (t_f)|K_0^{in}(t_i)>|^2
=|<K_0^{out} (t_f)|\bar{K}_0^{in}(t_i)>|^2  ~.
\label{KK}
\end{equation}
Any deviation from the above equality will definitely signal
time-reversal
violation.
The comparison of the   probabilities of a
$\bar{K}^0$ transforming into $K^{0}$,
and  $K^0$ transforming into $\bar{K}^{0}$
can demonstrate  a  departure from time-reversal invariance.
More explicitly, such a departure is manifest in the
asymmetry
\begin{eqnarray}
A_T &=& \frac
{ P_{\bar{K} K} (\Delta t) - P_{K \bar{K}} (\Delta t)
}
{ P_{\bar{K} K} (\Delta t) + P_{K \bar{K}} (\Delta t)
} ~,
\nonumber \\
&=&
\frac{
|<K_0^{out}(t_f)|\bar{K}_0^{in}(t_i)>|^2
-|<\bar{K}_0^{out} (t_f)|K_0^{in}(t_i)>|^2
} {
|<K_0^{out}(t_f)|\bar{K}_0^{in}(t_i)>|^2
+|<\bar{K}_0^{out} (t_f)|K_0^{in}(t_i)>|^2
}
 ~, \label{Kabir}
\end{eqnarray}
known in the literature as the Kabir asymmetry \cite{Kabir}.

The time evolution from $t_i$ to $t_f$ is induced by the effective
Hamiltonian $H$:
\bea
A_{K_0 \rightarrow \bar{K}_0}
& = &
<\bar{K}_0^{out} (t_f)|K_0^{in}(t_i)>
{}~~=~~<\bar{K}_0^{out}|e^{-iH\Delta t}|K_0^{in}> ~,
\nonumber \\
A_{\bar{K}_0 \rightarrow K_0}
& = &
<K_0^{out} (t_f)|\bar{K}_0^{in}(t_i)>
{}~~=~~<K_0^{out}|e^{-iH\Delta t}|\bar{K}_0^{in}> ~.
\eea

Inserting the unity operator
\bea
{\bf 1} = |K_L^{in}><K_L^{out}| + |K_S^{in}> <K_S^{out}| ~,
\eea
to the right
of  the evolution operator $e^{-iH\Delta t}$ and using the fact that
$K_{L,S}$ are Hamiltonian eigenstates,
we obtain:
\bea
A_{K_0 \rightarrow \bar{K}_0}
& = &
<\bar{K}_0^{out}|K_L^{in}> <K_L^{out}|K_0^{in}>  e^{-i \lambda_L
\Delta t}  \nonumber \\
& + &
<\bar{K}_0^{out}|K_S^{in}> <K_S^{out}|K_0^{in}>  e^{-i \lambda_S
\Delta t}   \nonumber \\
& =  &
\frac{1}{N^2} ~ (1-\a) (1-\b) ~
( e^{-i \lambda_S \Delta t} - e^{-i \lambda_L \Delta t} ) ~,
\eea
and
\bea
A_{\bar{K}_0 \rightarrow K_0}
& = &
<K_0^{out}|K_L^{in}> <K_L^{out}|\bar{K}_0^{in}>  e^{-i \lambda_L
\Delta t} \nonumber \\
& + &
<K_0^{out}|K_S^{in}> <K_S^{out}|\bar{K}_0^{in}>  e^{-i \lambda_S
\Delta t}   \nonumber \\
& = &
\frac{1}{N^2} ~ (1+\a) (1+\b) ~
( e^{-i \lambda_S \Delta t} - e^{-i \lambda_L \Delta t} ) ~.
\eea

We see  that the time-dependent factor
$ g(\Delta t)  \equiv
( e^{-i \lambda_S \Delta t} - e^{-i \lambda_L \Delta t} )$, whose
absolute value square is given by
\bea
|g(\Delta t)|^2 =
e^{- \Gamma_S \Delta t} +
e^{- \Gamma_L \Delta t} - 2 cos (m_L-m_S) ~\Delta t ~
e^{-\frac{ \Gamma_S + \Gamma_L}{ 2} \Delta t} ~,
\eea
is common in both amplitudes and
therefore will cancel in the asymmetry $A_T$, which
becomes {\it time-independent}\cite{Kabir}. Thus
\bea
A_T =
\frac{
| (1+\a) (1+\b) |^2 - | (1-\a) (1-\b) |^2}
{| (1+\a) (1+\b) |^2 + | (1-\a) (1-\b) |^2} ~,
\eea

Making the substitutions $\a = \epsilon + \delta$ and
$\b = \epsilon -\delta$, and keeping only linear
terms, one finds that
\bea
A_T \approx 4 Re \; [\epsilon] ~.
\eea
We note therefore that a non-zero value for $A_T$ signals a direct
measurement of
$T$-violation without any assumption about $CPT$ invariance.

To make clear the misunderstandings in the
literature,\cite{lee-wu}--\cite{Maiani} (with the exception of
ref.\cite{CORRECT})  we need to
introduce
the adjoint  outgoing states:
\bea
<K_S^{in}|&=&\frac{1}{N^*}\left((1+\alpha^*)<K_0^{in}| +(1-\alpha^*)
<{\bar K}_0^{in}| \right) ~,\nonumber\\
<K_L^{in}| &=& \frac{1}{N^*}\left((1+\beta^*)<K_0^{in}| -(1-\beta^*)
<{\bar K}_0^{in}|
\right) ~.
\eea

Notice that the adjoint states $<K_S^{in}|$ and $<K_L^{in}|$, are not
orthogonal
to $|K_S^{in}>$ and  $|K_L^{in}>$:
\bea
<K_S^{in}|K_S^{in}>&=& \frac{1+|\alpha|^2}{|1-\alpha\beta|} ~,~~~~~~
<K_L^{in}|K_L^{in}> ~=~ \frac{1+|\beta|^2}{|1-\alpha\beta|} ~,
\nonumber\\
<K_S^{in}|K_L^{in}>&=& \frac{\alpha^*+\beta}{|1-\alpha\beta|}
{}~,~~~~~~
<K_L^{in}|K_S^{in}> ~=~ \frac{\alpha+\beta^*}{|1-\alpha\beta|} ~,
\nonumber\\
\rightarrow ~~~
<K_S^{in}|K_L^{in}>&+&<K_L^{in}|K_S^{in}> ~=~
\frac{2 Re ~[(\alpha+\beta)]}{|1-\alpha\beta|} ~~=~~
\frac{4Re~ [\epsilon]}{|1-\alpha\beta|} ~.
\eea

In linear order in $\epsilon$ and $\delta$, the
approximate equality
\bea
A_T ~ \approx ~ <K_S^{in}|K_L^{in}>+<K_L^{in}|K_S^{in}>
{}~ \approx ~ 4 Re~[\epsilon] ~,
\label{Tviol}
\eea
holds. This relation resulted in
some misleading conclusion in the literature, namely that
$A_T \neq 0$ is not associated with $T$-violation,
but rather with the non-orthogonality of the physical
incoming states $K_L^{in}$
and $K_S^{in}$ states, and with the violation
of $CP$. However, as we already stressed, (i) the
relevant physical states
$<K_L^{out}|$ and $|K_S^{in}>$
are {\it always orthogonal} (see eq. (\ref{ortho} )
 and (ii) $A_T$ is  {\it by definition} the magnitude of
$T$-violation,
without any assumption about the validity of $CPT$ or even unitarity.

To better illustrate the misunderstanding,
let us imagine
that the $CP$-violating part
$\b$ of $K_L$ is zero. In this case $\epsilon = - \delta$, so
that $T$ is violated together
with $CPT$, with $CP$ invariance  in the $K_L$ decays.
Besides, if $CPT$ is assumed, then $\delta=0$ and
$\epsilon= \a =\b$.
In that case, clearly, $T$-violation is identical to $CP$-violation.

\section{The CPLEAR measurement}

Up to now, we described the behaviour of the
theoretical asymmetry that
stems directly from the definition of
$T$-reversal. However, as we mentioned
in the introduction, CPLEAR uses semi-leptonic
decays in order to tag the strangeness of the
final states and therefore the experimental
asymmetry of eq.(\ref{at1}) is:
\bea
A_T^{exp} =
\frac{\overline{R}_+ ~ (\Delta t) - R_{-} ~(\Delta t)}
{\overline{R}_+ ~(\Delta t) + R_{-} ~(\Delta t)}
\label{at2} ~,
\eea
where
\bea
\overline{R}_+
{}~(\Delta t)
& = &
| <e^+ \pi^- \nu(t_f)~ |K_0^{in}(t_f)>
<K_0^{out}(t_f)~ |\bar{K}_0^{in}(t_i)> |^2 ~,\nonumber \\
R_{-} ~(\Delta t)& = &
| <e^- \pi^+ \bar{\nu}(t_f)~ |\bar{K}_0^{in}(t_f)>
<\bar{K}_0^{out}(t_f)~ |K_0^{in}(t_i)> |^2 ~.
\eea

The basic idea here is the following:
There are in principle four semi-leptonic decays for neutral kaons:
\bea
K^0 & \rightarrow  &  e^+ \pi^- \nu ~, ~~~
\bar{K}^0  \rightarrow    e^- \pi^+ \bar{\nu} ~,\nonumber \\
K^0 & \rightarrow  &  e^- \pi^+ \bar{\nu} ~, ~~~
\bar{K}^0  \rightarrow    e^+ \pi^- \nu ~.
\eea
Among them, the first two are characterized by
$\Delta S = \Delta Q$ and are
allowed, while the others
are characterized by
$\Delta S = -\Delta Q$ and would be forbidden
if no oscillations between
$K^0$ and $\bar{K}^0$ were occurring. By looking
therefore at the ``wrong-sign'' leptons, one
studies $K^0-\bar{K}^0$ conversions.

As we see from the above expressions, the squared matrix elements
\bea
| <e^+ \pi^- \nu(t_f)~ |K_0 ^{in}(t_f)> |^{2}
& \equiv & |a|^2 ~|1-y|^2 ~,\nonumber \\
| <e^- \pi^+ \bar{\nu}(t_f)~ |\bar{K}_0^{in}(t_f)> |^2
& \equiv & |a|^2 ~|1+y|^2 ~,
\eea
enter in the calculation
and are parametrized by the
quantity $y$ \cite{rev,Maiani}, which describes
$CPT$-violation in semileptonic decays,
when the $\Delta S = \Delta Q$ rule holds.
Moreover, although the
$\Delta S = \Delta Q$ rule is expected from
the Standard Model to be valid up to
order $10^{-14}$ \cite{rev}, the experimental
limit before CPLEAR was much larger \cite{pdg}. For this
reason, two quantities (denoted by $x$ and $\bar{x}$
\cite{rev,Maiani}),
which are
 related to violation of
the $\Delta S = \Delta Q$ in the decays,
have been retained in the analysis \cite{at_paper}.
These parameters were
found to be very small, and will not
concern us further.

Even if $y$ is included in the calculation the time-independence of
the asymmetry still holds. However, $y$ does enter in the
asymmetry calculation:
\bea
A_T =
\frac{| (1+\a) (1+\b) |^2|1-y|^2 - | (1-\a) (1-\b) |^2|1+y|^2}
{| (1+\a) (1+\b) |^2|1-y|^2 + | (1-\a) (1-\b) |^2|1+y|^2} ~.
\eea
In particular for the
linear approximation one finds that
\bea
A_T^{exp} \approx 4 Re \; [\epsilon] ~ - 2 Re ~ [y] ~.
\eea
Since $y$ has also been measured by the experiment and
is found to be close to zero \cite{NEWCPLEAR}, we
conclude that the non-zero value
of $A_T^{exp}$ is due to $T$-violation.

One basic point to emphasize here, is
that CPLEAR
uses only one out of the possible decaying channels,
and therefore  its measurements
are independent of {\it any unitarity assumption}
and the possible existence of invisible decay modes.
An interesting question to ask, however,
is what information one could obtain from previous measurements
plus unitarity \cite{bell-stein,phas,lee-book,NEWCPLEAR}. Unitarity
implies the relations
\bea
<K_L^{in} |K_S^{in}> & = & \Sigma_{f}
<K_L^{in} |f^{in}> <f^{out} |K_S^{in}> ~,
\nonumber \\
<K_S^{in} |K_L^{in}> & = & \Sigma_{f}
<K_S^{in} |f^{in}> <f^{out}| K_L> ~,
\eea
where $f$ stands for {\it all}
possible decay channels.
Making the additional assumption
that the final decay modes satisfy the relation
\, $|f^{in}> = |f^{out}> \equiv <f^{out}|^\dagger$ \,
(which is equivalent to making use of
$CPT$-invariance of the final state interactions), it is
possible to calculate the sum
$ <K_L^{in}| K_S^{in}> + <K_S^{in} |K_L^{in}> $,
by {\it measuring only the branching ratios of kaon decays}.
This is what is done in $K_L$, $K_S$ experiments, where only
the {\it incoming kaon states} are used. In the linear approximation,
this sum is equal
to $4 ~Re~[\epsilon]$
(see eq. (\ref{Tviol})).
However, this is an {\it indirect}
determination of $T$-violation, and
would not have been possible if invisible decays or
$CPT$-violation in the final states interactions
were present\cite{ehns}--\cite{John}. This is to be contrasted with
the results of CPLEAR, which  do not rely at
all on unitarity and thus on the knowledge
of other decay channels than the one used
in the analysis.

\section{Concluding comments}

Motivated by the recent CPLEAR report on the
first direct observation of time-reversal non-invariance,
we attempted to clarify
the situation on measurements of charge conjugation,
parity violation and time reversibility,
in systems with non-Hermitean Hamiltonians.
To do so, we re-discussed  the formalism of the neutral kaon system,
paying particular attention in the definition
of states in the vector space of the system,
but also in its dual and in the dual complex spaces.
This allows a consistent implementation
of the orthogonality conditions for the incoming and outgoing states,
used to describe particle-antiparticle
mixing and the time evolution of the system.

As a result, we confirm that the asymmetry measured
by CPLEAR, is directly related to the definition
of $T$-violation. In addition, it
does not get affected by time and decay
processes.  Finally, the experiment
uses only one out of the possible decaying channels,
therefore  its results
are independent of any $CPT$ or unitarity assumption,
and the possible existence of invisible decay modes.
We conclude therefore that, CPLEAR indeed made
the first direct measurement of $T$-violation.

\vspace*{0.4 cm}

{\bf \large Acknowledgements: }
We would like to thank the CPLEAR Collaboration
and in particular M. Fidecaro and P. Kokkas, for many fruitful
discussions on the CPLEAR results; we also thank A. Kehagias,
for illuminating discussions and comments. The work of C.K.
has been partially supported by the TMR contract
ERB-4061-PL-95-0789.


\end{document}